\documentclass{article}

\usepackage[preprint]{style}

\usepackage[utf8]{inputenc} 
\usepackage[T1]{fontenc}    
\usepackage{hyperref}       
\usepackage{url}            
\usepackage{booktabs}       
\usepackage{amsfonts}       
\usepackage{nicefrac}       
\usepackage{microtype}      
\usepackage[usenames,dvipsnames]{xcolor}
\usepackage{amsthm, amsmath}
\usepackage[linesnumbered,ruled,vlined]{algorithm2e}

\title{Looking for Fairness in Recommender Systems}

\author{%
  Cécile Logé \\
  Department of Computer Science \\
  Stanford University \\
  \texttt{ceciloge@stanford.edu} \\
}

\usepackage{tcolorbox}
\usepackage{amssymb}
\tcbuselibrary{theorems}
\newtcbtheorem
  []
  {definition}
  {Definition}
  {%
    colback=CadetBlue!5,
    colframe=CadetBlue!45!black,
    fonttitle=\bfseries,
  }
  {def}

\newtcbtheorem
  []
  {proposal}
  {Proposal}
  {%
    colback=CadetBlue!5,
    colframe=CadetBlue!45!black,
    fonttitle=\bfseries,
  }
  {def}

\begin{document}

\maketitle

\section{Introduction}
\label{intro}
Recommender systems can be found everywhere today, shaping our everyday experience whenever we’re consuming content, ordering food, buying groceries online, or even just reading the news. Let’s imagine we’re in the process of building a recommender system to make content suggestions to users on social media. When thinking about fairness, it becomes clear there are several perspectives to consider: the users asking for tailored suggestions, the content creators hoping for some limelight, and society at large, navigating the repercussions of algorithmic recommendations.

A shared fairness concern across all three is the emergence of  filter bubbles \citep{10.5555/2361740, areeb2023filter} , a side-effect that takes place when recommender systems are almost "too good", making recommendations so tailored that users become inadvertently confined to a narrow set of opinions/themes and isolated from alternative ideas.  From the user's perspective, this is akin to manipulation. From the small content creator's perspective, this is an obstacle preventing them access to a whole range of potential fans. From society's perspective, the potential consequences are far-reaching, influencing collective opinions, social behavior and political decisions.

How can our recommender system be fine-tuned to avoid the creation of filter bubbles, and ensure a more inclusive and diverse content landscape? Approaching this problem involves defining one (or more) performance metric to represent diversity, and tweaking our recommender system's performance through the lens of fairness. By incorporating this metric into our evaluation framework, we aim to strike a balance between personalized recommendations and the broader societal goal of fostering rich and varied cultures and points of view.

\section{Definitions}
\label{recommender}
\subsection{Mathematical Model}
A recommender system can be thought of as a two-step model that first (step 1) will predict a user's ratings on different items of content they haven't consumed yet, then (step 2) will surface a ranked list of top $k$ items of content to recommend to that user (based on the predicted ratings). 

\paragraph{Step 1 | Rating Prediction:} A typical approach for (step 1) is Collaborative Filtering (CF), a technique that makes predictions about the potential interests of a single user by collecting the likes and dislikes of many other users \citep{su2009survey}.  More specifically, the concept of Matrix Factorization (MF) gained "State of the Art" status during the 2007 Netflix Prize competition when the first prize team \citep{10.1145/1281192.1281206, 5197422} won with a model predicting individual users' movie ratings through latent features determined during training. 

We'll base our approach on MF, and we'll consider:
\begin{itemize}
    \item A set of $n$ users $U = \{u_1, ..., u_n\}$, and a set of $m$ items of content $C = \{c_1, ..., c_m\}$,
    \item A sparse matrix $R \in \mathbb{R}^{n\times m}$ containing \textit{actual} user ratings, meaning that $r_{ij}$ is $u_i$'s rating over $c_j$, and a complete matrix $\hat{R} \in \mathbb{R}^{n\times m}$ containing \textit{predicted} user ratings. 
\end{itemize}
Note that while $R$ is sparse, we'll assume every user/piece of content has at least one rating, meaning there is no empty row or column. We'll also assume ratings go from $1$ to $5$, although the model can be adjusted to work on a binary type of rating (e.g. like or dislike) as well. \\
To obtain the prediction matrix $\hat{R}$, we will assume the existence of $d$ latent features such that our rating matrix can be represented as the product of two matrices $Q \in \mathbb{R}^{n \times d}$ and $P\in \mathbb{R}^{d \times m}$. We can then think of $u_i$'s predicted rating over $c_j$ as:

    \begin{align*}
            \hat{r}_{ij} = q_i^Tp_j 
    \end{align*}
where $q_i$ and $p_j$ are vector columns, respectively representing $Q$'s $i$th row and $P$'s $j$th column.  

From there, the objective becomes to estimate the latent features, in other words, find $Q$ and $P$. A typical optimization goal for CF is to minimize RMSE (Root Mean Squared Error), so we can use the following loss function:
    \begin{align*}
            L(P, Q) &=\|R-\hat{R}\|_{\mathrm{F}}+\alpha\|Q\|_{\mathrm{F}}+\beta\|P\|_{\mathrm{F}} \\
            &=\sum_{j=1}^m \sum_{i=1}^n\left(r_{ij}-\hat{r}_{ij}\right)^2+\alpha \sum_{i=1}^n\left\|q_i\right\|_2+\beta \sum_{j=1}^m\left\|p_j\right\|_2
    \end{align*}

where ${\mathrm{F}}$ represents the Frobenius norm and, $\alpha\|Q\|_{\mathrm{F}}$ and $\beta\|P\|_{\mathrm{F}}$ are regularization terms for both matrices $Q$ and $P$. Note that because $R$ is sparse, the loss function should only ever look at non-empty values in $R$ and the corresponding predictions in $\hat{R}$. One interesting way to optimize this function is the Alternating Square Method \citep{logebuilding}, but we will refrain from going into the details as they are not the main focus of this study. 

\paragraph{Step 2 | Item Ranking:} Once our model is trained, we can rank the items of content $c_j$ for each user $u_i$ from highest to lowest rating $\hat{r}_{ij}$ and recommend the top $k$. Specifically, we'll refer to the set of indexes of the $k$ pieces of content recommended to user $i$ as $\texttt{top}(i, k)$.

To measure the relevance and quality of a ranking in recommender systems,  it is typical to use a metric such as \textbf{nDCG} (Normalized Discounted Cumulative Gain), where we look at the discounted value of the obtained ranking versus the discounted value of the ideal (or \textit{true}) ranking for each individual user.
Other classic metrics include \textbf{Accuracy@k}, \textbf{Precision@k}, \textbf{Recall@k}. Note that none of these metrics, as typically defined, measure for diversity or fairness. 

\subsection{Fairness Notions}
\paragraph{Concerns:} From the user's standpoint, the assumption behind CF is that users who liked similar content in the past will share similar preferences in the future. This assumption contributes to the creation of filter bubbles \citep{10.5555/2361740, areeb2023filter} where users end up being predominantly exposed to content that aligns with their existing preferences and ultimately get limited exposure to diverse content. \\
Even more alarming is the idea that CF can tighten these filter bubbles over time, with recommendations gradually becoming less diverse and sometimes even more radical. Imagine a user being consistently showed content from a particular political perspective, and slowly becoming isolated from alternative viewpoints. Being fed the same viewpoints over and over again, this user starts to engage more with them, clicking, commenting, maybe even liking some of the posts, thus sending "positive" signals to the model. The model will then gradually cluster this user with others who like this political perspective and perhaps even more radical ones, thus creating even more isolation and radicalization. And so on. Indeed, \citet{10.1145/2566486.2568012} found that recommender systems in general - and CF techniques in particular - tend to expose users to a slightly narrowing set of items over time. 

From the content creator's standpoint, this poses an issue as well. Because our model relies on past user-content interactions, it is likely to recommend popular content more frequently as these will have received more user interactions, thus coming up more often and more positively during training across a wide range of users \citep{fleder2009blockbuster}. This can inadvertently trigger a vicious cycle of under-representation and neglect for less popular, newer or more niche items of content.

\paragraph{Diversity:} The concept of \textbf{diversity} addresses these concerns from both standpoints. In the rest of our study, we will, however, focus on the user's side.\\ 
In their work on diversity for recommendation systems, \citet{antikacioglu2019new} introduced the concept of \textbf{ILD} (Intra-List Distance), to calculate the average pairwise distance among the items of a given list and measure the diversity of the recommendations sent to each user. As it is defined, and for the purpose of our work, the metric brings some limitations:
\begin{itemize}
    \item The use of absolute distances make the metric difficult to interpret. A normalized version of ILD (similar to nDCG) would be much more insightful and allow for overall goal-setting as well as comparisons between recommenders.
    \item With MF, all we have to begin with are past ratings from users $u_i$ on pieces of content $c_j$, which limits the way we can think about the notion of distance, whether it is between users or between pieces of content.
\end{itemize}

To fit our purposes as well as counter the challenges of MF, we propose our own normalized version of ILD called \textbf{nILLD} (normalized Intra-List Latent Distance) and suggest using it to evaluate our MF recommender \textit{after training}, specifically drawing on \textit{latent variables} from the matrix  $P \in \mathbb{R}^{d \times m}$ as a proxy for content feature vectors. 
\begin{definition}{$\texttt{nILLD}$}{nilld}
For a given distance metric $\texttt{dist}: \mathbb{R}^d \times \mathbb{R}^d \to \mathbb{R}_{\geq 0}$, we define the normalized Intra-List Latent Distance $\texttt{nILLD}(i, k, \texttt{dist})$ as the normalized sum of pair-wise distances between the $k$ recommended items selected for a given user $i$:
    \begin{align*}  
            \texttt{nILLD}(i, k, \texttt{dist}) &= \frac{1}{\texttt{maxILLD}(k, \texttt{dist})} \sum_{(j, j') \in \texttt{top}(i, k)} \texttt{dist}(p_j, p_{j'}) 
    \end{align*}
where $p_j$ represents the $j$th content vector from the recommender's latent matrix $P \in \mathbb{R}^{d \times m}$, and $\texttt{maxILLD}(k, \texttt{dist})$ is the maximum sum of pair-wise distances possible obtained for any subset of $C$ of size $k$: 
    \begin{align*}  
            \texttt{maxILLD}(k, \texttt{dist}) &= \max_{C^{(k)}} \sum_{(j, j') \in C^{(k)}} \texttt{dist}(p_j, p_{j'}) \\
    \end{align*}

From there, we also define the overall $\texttt{nILLD}(k, \texttt{dist})$ of a given recommender as the average normalized Intra-List Latent Distance across all $n$ users, for recommendation lists of size $k$:
    \begin{align*}  
            \texttt{nILLD}(k, \texttt{dist}) &= \frac{1}{n} \sum_{i=1}^n \texttt{nILLD}(i, k, \texttt{dist}) 
    \end{align*}
\end{definition}

As a result, we obtain a metric ranging from $0$ to $1$, with $\texttt{nILLD}(k, \texttt{dist}) =1$ corresponding to the maximum diversity possible. Immediately, we can see that there will be a trade-off between efficiency (measured with the classic performance metrics to assess of the relevance of the recommendations made to the user) and fairness (measured with our new diversity metric).

\begin{definition}{$\alpha$-Diversity}{diverse}
We say that a recommender system satisfies $\alpha$-diversity at $k$ with regards to a given distance metric $\texttt{dist}: \mathbb{R}^d \times \mathbb{R}^d \to \mathbb{R}_{\geq 0}$ if: 
    \begin{align*}  
            \texttt{nILLD}(k, \texttt{dist}) \ge \alpha 
    \end{align*}
We say that a recommender system satisfies \textit{individual} $\alpha$-diversity at $k$ with regards to a given distance metric $\texttt{dist}: \mathbb{R}^d \times \mathbb{R}^d \to \mathbb{R}_{\geq 0}$ if: 
    \begin{align*}  
            \forall i = 1, ..., n, \texttt{  nILLD}(i, k, \texttt{dist}) \ge \alpha 
    \end{align*}
\end{definition}

The idea behind $\alpha$-diversity is to assess the recommender system's ability to maintain a minimum level of diversity. However, it is clear from the definitions that $\alpha$-diversity is a rather weak notion as it only gives information about the recommender as a whole. Even if we were to measure $\alpha$-diversity on different groups of users, it would still come with the usual pitfalls associated with group fairness. Instead, we prefer to focus on individual $\alpha$-diversity, a notion closer to individual fairness in essence. The idea is to guarantee each individual user is offered a set of recommendations that is deemed varied enough to avoid falling into the traps of filter bubbles. 

\subsection{Proposal for a Post-Processing Algorithm}
While our new diversity metric is useful for assessing and comparing recommender systems through the lens of diversity, one caveat is that it comes in \textit{after the fact}, once our recommender is fully trained and once the top $k$ recommendations have been selected for each user. Thus it is not obvious that this metric can help \textit{improve} a recommender system on its own. \\
 
Complementary to $\alpha$-diversity, we introduce an algorithmic post-processing step to "correct" the recommendations made by our recommender whenever necessary, keeping in mind that recommender systems need to strike a balance between providing personalized and diverse recommendations. Therefore, the goal of this step is to achieve individual $\alpha$-diversity while maintaining relevance for the user as much as possible, meaning impacting the standard performance metrics (such as nDGC) as little as possible. Concretely, once a recommender system has been trained (step 1) and during the selection of $k$ items to be recommended for each user $i$ (step 2), we propose to run Algorithm 1: \\ 
\begin{algorithm}
\caption{Post-Processing for Individual $\alpha$-Diversity}
\KwData{Recommender output i.e. top $k$ recommendations for each user i}
\KwResult{Updated recommendations to satisfy individual $\alpha$-diversity}
\ForEach{user $i$}{    
    $\texttt{nILLD}_i \leftarrow \texttt{nILLD}(i, k, \texttt{dist})$\;    
    $\texttt{r}_i \leftarrow \min (\hat{r}_{ij})$ over $j$ in $\texttt{top}(i, k)$\;
    \While{$\texttt{nILLD}_i < \alpha$}{        
        $j \leftarrow \arg\min \sum_{j' \in \texttt{top}(i, k)} \texttt{dist}(p_j, p_{j'})$ over $j$ in $\texttt{top}(i, k)$\;
        Remove $j$ from $\texttt{top}(i, k)$\;
        $j'' \leftarrow \arg\max \sum_{j' \in \texttt{top}(i, k)} \texttt{dist}(p_{j''}, p_{j'})$ over $j''$ \textbf{not} in $\texttt{top}(i, k)$ s.t. $\hat{r}_{ij''} \geq \texttt{r}_i$\;
        Add $j''$ to $\texttt{top}(i, k)$\;

        $\texttt{nILLD}_i \leftarrow \texttt{nILLD}(i, k, \texttt{dist})$\;    
        \If{$j'' = j$}{$\texttt{r}_i \leftarrow \texttt{discount}*\texttt{r}_i$\;}
    }
}
\end{algorithm}
\\ As long as the nILLD for user $i$ remains below $\alpha$, our proposed algorithm will iteratively replace the least diverse recommendation from the list with a more diverse one while guaranteeing a minimum acceptable rating $\texttt{r}_i$, initially set as the minimum predicted rating among the top $k$ recommendations. The process continues - and progressively adjusts $\texttt{r}_i$ down at a chosen $\texttt{discount}$ rate - until the diversity threshold is met, thus ensuring a set of diverse but still relevant recommendations for each user.

\section{Related Work}
\label{related}
Several other works have considered the issue of diversity in recommender systems \citep{zhao2024fairness}, although few have truly grounded their frameworks in notions of fairness, instead choosing to focus on solving for more efficiency, often from a purely commercial standpoint. \\
Our new diversity metric is directly inspired from the work of \citet{antikacioglu2019new}. However, their approach focuses heavily on diversifying \textit{categories} of items and \textit{types} of users, which can be seen as too wide when it comes to fairness - and certainly too weak in the case of individual fairness. In contrast, \citet{10.1145/2348283.2348310} propose a similar approach to ours, using latent features (e.g. from an MF model) to create a Latent Factor Portfolio and cater to individual users. More specifically, their idea is to capture the user's interest range (in other words, their appetite for diversity) and adjust recommendations accordingly. However, while their motivation stems from improving the user's experience, it does not address fairness directly. In fact, their approach would likely go right past the issues related to filter bubbles, as the user's interest range is implicitly measured from the data, meaning from past user-item interactions; What would the data say of a user stuck in a filter bubble and only engaging with a limited range of content? Is it fair to not offer diverse content to a user who hasn't implicitly signaled for it? \\
Other teams have also proposed post-processing steps to correct for diversity in recommender systems during the ranking stage. For example, \citet{10.1145/1060745.1060754} introduce a Topic Diversification Algorithm aimed at adjusting final recommendation lists to make them more diverse and prevent the "portfolio effect" - when a user systematically gets recommended items from, say, the same author or the same website. But again, the main goal is not to obtain a fairer system for the user, as much as it is to prevent "the law of diminishing marginal returns" where users are less prone to spend money on a product after having been recommended similar ones too many times before.\\
If filter bubbles are a concern in recommender systems, and we believe they are \citep{areeb2023filter}, then diversity needs to be directly tied to notions of fairness instead of being used to achieve more efficiency and increase clicks and returns. Users need to be kept at the front and center of the approach, meaning diversity needs to be measured at an individual level (e.g. with individual $\alpha$-diversity), and we need a way to enforce it (e.g. with a post-processing step). The main goal of our work is to be able to guarantee a minimum level of diversity for each user while maintaining overall performance and satisfaction.

\section{Next Steps}
\label{next}

While our diversity framework is now properly defined, we have yet to prove the relevance and feasibility of these definitions in theory and in practice.

\paragraph{Diversity:} Proving the relevance of our new metrics and determining the optimal value for the diversity threshold $\alpha$ in our proposed post-processing algorithm will be crucial. Indeed, $\alpha$ needs to be high enough to align with the overarching goal of promoting fairness and preventing the formation of filter bubbles, but it shouldn't be so high that recommendations are no longer personalized to each user. 
We would like to conduct a thoughtful exploration including:
\begin{enumerate}
    \item Building vanilla recommender systems with matrix factorization (e.g. using the MovieLens 20M dataset, the Netflix dataset, the Yahoo! Front Page Click Log) and measuring overall as well as individual $\texttt{nILLD}$s (along with classic statistics like min, max, std). This will serve as a first baseline. 
    \item Running sensitivity analyses by varying $\alpha$, and measuring the impact of enforcing individual $\alpha$-diversity on classic performance metrics, so as to assess the size of the trade-off between fairness/diversity and relevance (e.g. nDCG), but also on the actual diversity of recommendations using (a) actual item features from the datasets that provide them, (b) manual exploration of the different recommendation lists for a random set of users.
    \item Conducting the same analysis in recommender systems that are known to promote the creation of filter bubbles - if possible - so as to obtain an idea of what \textit{not enough diversity} means in terms of $\texttt{nILLD}$ and $\alpha$, 
\end{enumerate}
\paragraph{Post-Processing:} While our first proposal makes sense, it must be tested further in real settings to assess efficacy and feasibility. In particular, reviewing and improving our algorithm in terms of computational complexity will be essential if we want it to be used in practice: in its current state, depending on the size of the dataset and the complexity of the distance metric, it introduces clear computational overhead due to its many iterations, potentially affecting scalability.

Ultimately, we hope this brings us to a better understanding of how filter bubbles are formed and how diversity can help improve these recommender algorithms for all users (and in turns make our experiences on the internet a little fairer and richer).

\textbf{Video URL:} https://drive.google.com/file/d/1E0\_9yElYv3GoVwE6AgqdglnpPVIcGhYF/

\bibliographystyle{rusnat}
\bibliography{references}

\begin{thebibliography}{12}
\providecommand{\natexlab}[1]{#1}
\providecommand{\EM}{\em}
\providecommand{\RNtxt}{\relax}
\RNtxt{}

\bibitem[Antikacioglu et~al.(2019)A.~Antikacioglu, T.~Bajpai, R.~Ravi]{antikacioglu2019new}
{\EM Antikacioglu Arda, Bajpai Tanvi, Ravi R.}
\newblock A new system-wide diversity measure for recommendations with efficient algorithms. 2019.

\bibitem[Areeb et~al.(2023)Q.~M. Areeb, M.~Nadeem, S.~S. Sohail, R.~Imam, F.~Doctor, Y.~Himeur, A.~Hussain, A.~Amira]{areeb2023filter}
{\EM Areeb Qazi~Mohammad, Nadeem Mohammad, Sohail Shahab~Saquib, Imam Raza, Doctor Faiyaz, Himeur Yassine, Hussain Amir, Amira Abbes}.
\newblock Filter Bubbles in Recommender Systems: Fact or Fallacy -- A Systematic Review. 2023.

\bibitem[Bell et~al.(2007)R.~Bell, Y.~Koren, C.~Volinsky]{10.1145/1281192.1281206}
{\EM Bell Robert, Koren Yehuda, Volinsky Chris}.
\newblock Modeling relationships at multiple scales to improve accuracy of large recommender systems \allowbreak\newblock// Proceedings of the 13th ACM SIGKDD International Conference on Knowledge Discovery and Data Mining. New York, NY, USA: Association for Computing Machinery, 2007.  95–104.
\newblock (KDD '07).

\bibitem[Fleder, Hosanagar(2009)D.~Fleder, K.~Hosanagar]{fleder2009blockbuster}
{\EM Fleder Daniel, Hosanagar Kartik}.
\newblock Blockbuster culture's next rise or fall: The impact of recommender systems on sales diversity \allowbreak\newblock// Management science. 2009. 55, 5. 697--712.

\bibitem[Koren et~al.(2009)Y.~Koren, R.~Bell, C.~Volinsky]{5197422}
{\EM Koren Yehuda, Bell Robert, Volinsky Chris}.
\newblock Matrix Factorization Techniques for Recommender Systems \allowbreak\newblock// Computer. 2009. 42, 8. 30--37.

\bibitem[Log{\'e}, Yoffe(2019)C.~Log{\'e}, A.~Yoffe]{logebuilding}
{\EM Log{\'e} C{\'e}cile, Yoffe Alexander}.
\newblock Building the optimal Book Recommender and measuring the role of Book Covers in predicting user ratings \allowbreak\newblock// Stanford Computer Science Department CS229. 2019.

\bibitem[Nguyen et~al.(2014)T.~T. Nguyen, P.-M. Hui, F.~M. Harper, L.~Terveen, J.~A. Konstan]{10.1145/2566486.2568012}
{\EM Nguyen Tien~T., Hui Pik-Mai, Harper F.~Maxwell, Terveen Loren, Konstan Joseph~A.}
\newblock Exploring the filter bubble: the effect of using recommender systems on content diversity \allowbreak\newblock// Proceedings of the 23rd International Conference on World Wide Web. New York, NY, USA: Association for Computing Machinery, 2014.  677–686.
\newblock (WWW '14).

\bibitem[Pariser(2012)E.~Pariser]{10.5555/2361740}
{\EM Pariser Eli}.
\newblock The Filter Bubble: How the New Personalized Web Is Changing What We Read and How We Think. USA: Penguin Books, 2012.

\bibitem[Shi et~al.(2012)Y.~Shi, X.~Zhao, J.~Wang, M.~Larson, A.~Hanjalic]{10.1145/2348283.2348310}
{\EM Shi Yue, Zhao Xiaoxue, Wang Jun, Larson Martha, Hanjalic Alan}.
\newblock Adaptive diversification of recommendation results via latent factor portfolio \allowbreak\newblock// Proceedings of the 35th International ACM SIGIR Conference on Research and Development in Information Retrieval. New York, NY, USA: Association for Computing Machinery, 2012.  175–184.
\newblock (SIGIR '12).

\bibitem[Su, Khoshgoftaar(2009)X.~Su, T.~M. Khoshgoftaar]{su2009survey}
{\EM Su~Xiaoyuan, Khoshgoftaar Taghi~M}.
\newblock A survey of collaborative filtering techniques \allowbreak\newblock// Advances in artificial intelligence. 2009. 2009.

\bibitem[Zhao et~al.(2024)Y.~Zhao, Y.~Wang, Y.~Liu, X.~Cheng, C.~Aggarwal, T.~Derr]{zhao2024fairness}
{\EM Zhao Yuying, Wang Yu, Liu Yunchao, Cheng Xueqi, Aggarwal Charu, Derr Tyler}.
\newblock Fairness and Diversity in Recommender Systems: A Survey. 2024.

\bibitem[Ziegler et~al.(2005)C.-N. Ziegler, S.~M. McNee, J.~A. Konstan, G.~Lausen]{10.1145/1060745.1060754}
{\EM Ziegler Cai-Nicolas, McNee Sean~M., Konstan Joseph~A., Lausen Georg}.
\newblock Improving recommendation lists through topic diversification \allowbreak\newblock// Proceedings of the 14th International Conference on World Wide Web. New York, NY, USA: Association for Computing Machinery, 2005.  22–32.
\newblock (WWW '05).

\end{thebibliography}


\end{document}